# Grid Security and Integration with Minimal Performance Degradation

Sugata Sanyal
School of Technology and Computer Science
Tata Institute of Fundamental Research, India
sanyal@tifr.res.in

Rangarajan A. Vasudevan
Department of Computer Science and Engineering
Indian Institute of Technology, Madras, India
ranga@cs.iitm.ernet.in

Ajith Abraham
Computer Science Department
Oklahoma State University, USA
ajith.abraham@ieee.org

Marcin Paprzycki
Computer Science Department
Oklahoma State University, USA
marcin@cs.okstate.edu

*Abstract*- **Computational grids are believed to be the ultimate framework to meet the growing computational needs of the scientific community. Here, the processing power of geographically distributed resources working under different ownerships, having their own access policy, cost structure and the likes, is logically coupled to make them perform as a unified resource. The continuous increase of availability of high-bandwidth communication as well as powerful computers built of low-cost components further enhance chances of computational grids becoming a reality. However, the question of grid security remains one of the important open research issues. Here, we present some novel ideas about how to implement grid security, without appreciable performance degradation in grids. A suitable alternative to the computationally expensive encryption is suggested, which uses a key for message authentication. Methods of secure transfer and exchange of the required key(s) are also discussed.**

*Keywords*- Grid computing, Security, Temporal distribution, Spatial distribution

## I. Introduction

Grid computing started in the 1990s as a response to the need for large amount of computing power. Scientific problems needed such computing power but (a) were bound by financial constraints that prevented investments in "supercomputers," and (b) it was realized that there exists an enormous pool of available and unused resources among computers connected across the internet. As a result, the concept of "coordinated resource sharing and problem solving among dynamic collections of individuals, institutions and resources" arose [2].

Currently, grid-related research is focused mostly on delivering the best available performance, including questions of load balancing, on the questions of resource discovery (which recently has been clouded due to some substantial changes in the underlying framework), grid-enabling existing legacy software and the likes. At the same time the question of grid security, while recognized as an important issue, remains somewhat on the backburner. There is a reasonable explanation for this situation. Security matters only if the computational infrastructure of the grid works well and effectively. Until the desired work is correctly distributed to the appropriate resources on the grid and the problem is efficiently solved and the results returned to the originator, there is no real reason to worry if the whole process can be done in a secure fashion.

Obviously, the research devoted to security has proceeded in the meanwhile, and currently the state of the art suggests the use of encryption techniques for security purposes [12] [14]. At the same time, Foster et al. [6] suggest to avoid the use of encryption on the grounds of exportability issues. While one can discuss the validity of the latter's arguments, it is obvious that the area of encryption-based solutions and encryption-oriented research is rather controversial from the point of view of its social implications [17]. Therefore, it is important to seek solutions that do not rely directly on cryptography. Furthermore, another issue that arises is that use of encryption requires a substantial amount of computational resources. This is particularly so since the size of the key has to increase to protect the encrypted data. This means that the computing power available on the grid, which is the primary reason for the grid's setup, is utilized to encrypt and decrypt information rather then to perform computations. This situation is particularly important when the grid is set up to complete a large number of short jobs rather then a very large job like the cancer research conducted by United Devices [16]. In this case, each of these smaller chunks has to be encrypted and decrypted and the portion of time used to perform these functions (instead of solving the problem) is relatively much larger than in the case when each job takes hours to complete. Therefore, in this paper, we propose less resource consuming methods of secure authentication and secure data transfer. For authentication, we discuss the idea of temporal distribution of key information. We also believe that Winnowing and Chaffing [3] provides a good substitute for encryption in terms of the secure data transfer.

The organization of this paper is as follows. A brief summary of research, carried out in the field of grid security, is presented in Section 2; then, we discuss our ideas about security methods for key exchange and offer a brief description of Winnowing and Chaffing, in Section 3 and finally, we conclude in Section 4.

## II. Related Research

Currently, the most renowned grid-related research is concentrated around the Globus project [1]. It consists of the work done by the Globus project researchers themselves and all the subsequent research that utilizes the Globus toolkit. Our brief introduction to the research activities in the area of grid computing is therefore based mainly on the work done under the Globus project umbrella. Foster et al. [2] offer an exposition of the "grid problem". They present a comprehensive introduction to the issues involved in grid computing. The Open Grid Services Architecture (OGSA) model is presented in [2] and [8]. A thorough overview of computational grids and technologies that existed at that time was presented in [9] though it is now outdated.

Coming now to the security aspect, [6] presents the security issues coexisting within a grid, and builds an architecture to address them. Some approaches to handle the issue of interoperability of local security solutions with global grid security policies are proposed in [7] and [10]. The topics of dynamic creation of services and trust domains, and diverse local mechanisms, as viewed from the angle of implementation for the Globus Project are presented in [11]. Data integrity is also one of the central concerns of large-scale distributed computing systems such as the grid, whose primary products are the results of computation. In order to maintain the integrity of this data, the system must be resilient to diverse attacks. Gilbert et al. proposes a trust-based model for grid participants based on use of reputation systems and associated feedback mechanisms [14]. Trust is essential especially to maintain the integrity of data being processed/produced in a grid environment. The problem of determining which resources to trust and distrust for the purposes of data integrity closely resembles some of the trust issues (reputation models) present in online auctions [22]. Online auctions like eBay have adopted such reputation systems to help cope with trust uncertainties inherent in Internet applications. Sarmenta addresses the grid security issues using a voting and spot-checking technique [15].

## III. Encryptionless Security

We begin this section with a technical comparison between encryption versus Winnowing and Chaffing (W&C) with an emphasis on the computational cost. Firstly, any encryption method like RSA and Advanced Encryption Standard (AES) does not guarantee data integrity and therefore it becomes essential to use a system of Message Authentication Codes (MAC) in addition to the encryption algorithm. Compared to this, W&C provides both data privacy and data integrity using MAC only. Hence, by a clear margin, W&C has smaller computational cost by avoiding the encryption completely. In the case when encryption is used with other forms of data integrity that are less computationally expensive than MAC – calculated using hashing functions – we will show that W&C is still cheaper than encryption. The only computationally intensive step in this method is the calculation of the MAC, which is done at the sender's and receiver's end. MACs are calculated using a combination of HMAC with a secure hashing algorithm like SHA-1 or MD-5. Let us now look at the cost involved in the AES encryption and in the HMAC-SHA1.

For a common data block size of 512-bits, calculation of the MAC using HMAC-SHA1 effectively requires around 700 of 32-bit XOR operations and 132 shifts of 32-bit words. Note that the maximum message size supported by SHA-1 is $2^{64}$, and the algorithm deals with messages in block increments of 512-bits. AES has no limit on the maximum data size and handles 128 message bits at a time. For a 128-bit message block, assuming the least key length of 128-bits, AES requires effectively nearly 400 32-bit XORs and 33 shift operations on 32-bit words. Including the multiplication and inverse operations, and scaling the above values for a message size of 512-bits, we find that AES effectively requires more than 1000 32-bit XORs, an equal number of 32-bit word shifts as W&C, 68 operations of calculating multiplicative inverses of 8-bit words and more than 350, 8-bit multiplications [18][19].

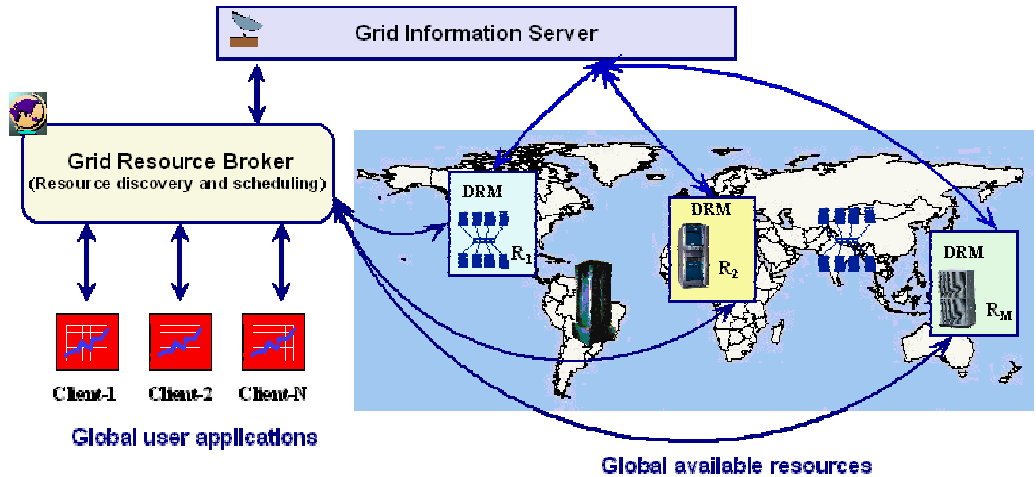

Figure 1. Computational grid model

For data sizes of lesser than 512-bits, the MAC algorithm pads the data to fit the 512-bit limit and then performs the operations. Padding is a low-cost operation that is comparative to a bit shift plus XOR operation. Thus across all data sizes, winnowing and chaffing is definitely cheaper than an encryption algorithm like AES. As for asymmetric key algorithms like RSA, the computational cost involved is more prohibitive than symmetric key encryption algorithms.

One big disadvantage of W&C is the increase in amount of transferred data due to the addition of chaff packets. To provide minimum security, for a "wheat" packet, at least one "chaff" packet is to be added. Therefore, for N blocks of "wheat", we need to transfer 2*N blocks. However, this does not translate to an analogous increase in the computational cost since, as specified in the algorithm, the MAC for the "chaff" packets are assigned randomly and not calculated. Therefore, the computational cost incurred is for the original N blocks of "wheat" data only, irrespective of the number of "chaff" blocks transferred. Also, a point to be noted is that the probability that a randomly generated MAC (which is 160-bits in length for the HMAC-SHA1) for a "chaff" packet is its correct MAC (thereby making the "chaff" look like "wheat" to the receiver) is $1/2^{160}$, which is practically negligible. A summary of the effective number of different operations performed in the course of each algorithm is presented below.

**MAC using HMAC-SHA1**
132 shift operations on 32-bit words; and, 762 XOR operations on 32-bit words.

**AES**
132 shift operations on 32-bit words; 1214 XOR operations on 32-bit words; 320 modulo multiplications in Finite Field of $2^8$ on 8-bit words; 44 multiplications of 8-bit words; and, 68 multiplicative inverse calculations in Finite Field of $2^8$ on 8-bit words.

While the computational nodes become constantly more powerful, the size of the key has to increase as well and thus the resource usage will remain relatively unchanged. Thus, a node may be constantly encrypting and decrypting data if it constantly receives requests and has to handle queries. One more fact, that is somewhat ironic, is that it is the very nature of distributed, grid-based computing that enables the breaking down of many encryption algorithms (for example, by a fast brute-force search across the complete space of available keys). Thus, minimal use of encryption would lessen computing time and reduce the risk of compromise-by-extensive-usage. Before we discuss our proposal for encryption-less security, we first present the model of the computational grid that we consider.

### A. Our Model of the Computational Grid

We consider the following computational grid model. A computational grid is created primarily to share the burden of solving a computational problem by accessing the computing power of its nodes. Figure 1 depicts the general framework for grid computing focusing on the interaction between Grid Resource Broker (GRB), Domain Resource

Manager (DRM) and the grid information server [20]. The Grid Resource Broker (GRB) is responsible for resource discovery, deciding allocation of a task (job) to a particular resource, binding of user applications (files), hardware resources, initiation of computations, adaptation to the changes in grid resources, and presentation of the grid to the user as a single, unified resource. It finally controls the physical allocation of the tasks and manages the available resources constantly while dynamically updating the grid scheduler whenever there is a change in resource availability. It is assumed that a resource or a group of resources having the program code and necessary data that together define the computational problem and coordinate the solution to the computational problem with the GRB. The DRMs with different capabilities that are networked to the GRB are dedicated to the grid. That is, these DRMs play a single role only, that of being members of that particular computational grid. As we will see later, this model has its advantages and disadvantages. The problem to be solved is split and concurrently solved at these dedicated DRMs in parallel. The result or results of execution of the program code on separate DRMs are collected by the GRB and organized into meaningful information that is used either as is or is kept for further processing.

The process that needs to execute at a dedicated DRM must first authenticate itself with that DRM. This authentication procedure should be secure to prevent entry to unauthorized programs. For this, we suggest the following methods, either one of which could be used, based on particular requirements. They all pertain to exchange of key information to secure subsequent communication and data transfer.

### B. Methods of Authentication and Key Transfer

The first method involves the only suggested use of encryption. Encrypting the authentication information and the secret key and transferring it to the receiver is a viable though costly option. This option facilitates dynamic generation of keys for communication with various nodes and at various times. However, as was elucidated before, this method is likely to be broken into in the case when the same set of authentication information is utilized over an extended period. The extended time gives adequate opportunities for an intruder to obtain the authentication information that is transmitted, and by using cryptanalysis combined with the power of the grid itself, the encryption could be broken. Thus, the information needed for authentication with a particular DRM should be differentiated across time. For example, a randomly generated key could be used but this key has to be conveyed to the DRMs beforehand, which creates a case of a vicious circle: to be able to securely send the key, we need to securely send the key first.

A second alternative to key exchange is the spatial split-distribution of key information. It is based on an extension of Shamir's secret sharing method [4] and utilizes the existence of multiple disjoint paths from the GRB to a DRM. In this method, the secret key $K$ is sent in the form of $N$ parts say $f_1, f_2, ..., f_N$ (as specified in what follows). First, let us consider a $(T, N)$-threshold scheme, where

a) knowledge of any $T$ or more pieces makes $K$ easily computable;

b) knowledge of any $T-1$ or lesser pieces leaves $K$ completely undetermined.

The scheme is designed as follows:

a) Construct a polynomial

$$f(x) = a_0 + a_1 x + a_2 x^2 + ... + a_{T-1} x^{T-1}$$

where $a_1, a_2, a_{T-1}$ are chosen randomly while $a_0$ is assigned to $K$.

b) Evaluate $f(x)$ at $N$ values $x_1, x_2, x_N$. That is, calculate $f(x_1), f(x_2), f(x_N)$.

Now given any $T$ of these $N$ values, it is possible to reconstruct $f(x)$ and hence obtain the value of the secret key $a_0$. These $N$ values are sent along $N$ mutually disjoint paths to the DRM so that the chances of an unauthorized intermediate DRM obtaining the secret key is minimal. The assumption behind this method is that it is practically infeasible for an intruder to monitor $T$ or more paths to obtain $T$ or more packets from a sender, simultaneously. This method also facilitates the distribution of randomly generated keys.

The third method takes a multimodal approach to security. In this system, the secret key could be agreed upon at the time of a contractual agreement between the parties that own a resource and the parties that run the processes (remember, that in our model, the DRMs belong 100% to the grid and are not used for any other purposes). The actual transfer of authentication information and data, however, could be carried out through the regular channels of communication in the grid. This might, however, mean a prior personal contact between a DRM that wants to participate in the grid and the group operating the grid, which might not always be possible. However, by following this multimodal approach, we can practically eliminate the chance of any compromise in confidential information. One disadvantage of this scheme is that every time the key needs to be changed, a meeting between the concerned parties should be arranged (thus making it susceptible to the brute-force cryptanalysis attacks if the key exchange – meeting – does not happen for an extended period of time).

It is possible to further modify the latter scheme by applying to it the temporal split-distribution of information. This facilitates dynamic generation of keys and at the same time benefits from the extended secrecy of the multimodal approach. The algorithm of the modified scheme is similar to the spatial split-distribution method presented earlier, but does not require the existence of multiple paths. In this algorithm, the information exchanged in person or by a contractual agreement is not the secret key. It is a prime number, $p$, use of which is essential to convey the key information in a secure manner, subsequently. The algorithm used by the sender for this method is as follows:

1. Split $K$ into $N$ arbitrary parts.

2. Assign the values of the bit positions where the splits took place to $R_1, R_2 ... R_{N-1}$.

3. Construct polynomial $P(x)$ using these $R_i$'s as roots.

4. Evaluate $P(x)$ at the $N$ different values $x_1, x_2, ... , x_N$ modulo the prime number p. Let us denote the pairs as $(x_i, P(x_i))$ for all $i$ from 1 to $N$.

5. Club $K_i$ along with the respective $(x_i, P(x_i))$ and send them across, one by one.

At the receiver's end, the algorithm for obtaining back the entire key $K$ is as follows:

1. Collect all the $(x_i, P(x_i))$ pairs from the packets and determine the polynomial $P(x)$ completely, since p is known.

2. Calculate the roots of $P(x)$. These give the position of the splits that took place at the sender's end.

3. Using this knowledge of the positions of the splits and the packet sequence number, obtain the correct sequence of bits from the respective packets. From this, the key $K$ can be obtained immediately.

This method is secure from key leakage because of two reasons. Firstly, the key $K$ is split arbitrarily and hence the data content in a packet that contains $K_i$, $(x_i, P(x_i))$ pair and some random bits cannot be determined easily. Secondly, the polynomial used by the sender cannot be constructed without knowing the prime $p$. When a new key is generated and it needs to be communicated, this method can be used. As can be seen, it is therefore possible to retain the advantage of dynamic generation of keys and at the same time benefit from the security provided by an exchange in person.

Thus, we have provided three methods and their subsequent modifications for the purposes of authentication and key exchange. This key subsequently helps in ensuring secure data transfer.

## C. Secure Data Transfer

For the issue of security in the transfer of data, an encryption-less method that still offers similar level of security, could be used. A good method that fits the bill is the Winnowing and Chaffing approach suggested by Rivest [3], which is summarized as below.

In this method, a Message Authentication Code (MAC) is added to every packet that is sent by the GRB. MAC is calculated using a standard algorithm like HMAC-SHA1 [5]. The parameters to this algorithm are the packet sequence number, the contents of the packet, and the secret key, which was exchanged earlier (see previous section). Once the grid DRM receives a packet, it first calculates the MAC itself and then checks whether it matches with the MAC sent with the packet. If so, it "knows" that the sender is the GRB, else it discards the packet as originating from a false source. In this way, data can be authenticated. Note that usage of a secret key for authentication is not "encryption" and does not involve computationally intensive operations with large operands.

Now, security is implemented on top of this message authentication by adding the so-called "chaff" packets. These are packets, which have the same format as the genuine data packets, but the MACs are deliberately set to the wrong value. On seeing a packet with a non-matching MAC, the grid DRM can promptly ignore it. However, any intruder monitoring traffic has no way of differentiating the right value of the MAC from the wrong, as he has no knowledge of the secret key. For some practical issues in the implementation of this idea, the reader is referred to [3].

The algorithm can be summarized as follows:

1. Calculate MAC using packet contents, packet

sequence number and secret key.

2. Create the "wheat" packets using the sequence number, contents and the MAC.

3. Create the "chaff" packets by randomly generating MAC numbers. For maximum security, the contents of the packet should be the inverted result of the contents of the "wheat" packet with the same sequence number.

At the receiver's end:

1. Calculate MAC using the packet contents, packet sequence number and secret key.

2. If the calculated MAC matched with MAC of the packet, then it is a "wheat" packet. Else, it is a "chaff" packet and can be thrown away.

Using winnowing and chaffing, data can be securely transmitted across the grid. There is minimal computing power waste on the part of the DRMs since all that is to be done is to calculate the MACs of the packets and accordingly perform a check. Similarly, there is minimal overhead for the GRB to calculate the MACs and generate the chaff. The former by a similar argument involves low computational time compared to encryption, while the latter can be done randomly with low computational cost again. There is no way a chaff packet can be distinguished from a wheat packet by anybody other than the base station and the grid node without knowledge of the secret key. Thus, we can obtain high security at lesser performance-degradation of the grid than when compared to the use of encryption algorithms.

**D.** *Discussion on the Computational Grid Model*

The model of the computational grid used in the previous section lends itself to the implementation of security we had illustrated earlier. This is because of the "dedicated" nature of the DRMs of the grid. By "dedicated", the implicit assumption is that the owner(s) of the grid, which in this case could be the GRB, are assured of the DRMs integrity. That is, it is assumed that a DRM would not participate in activities that are disruptive to the principles and functioning of the grid. This is important especially since data have to be processed at the DRMs and information returned to the GRBs. In the eventuality that the DRMs are indeed compromised, then, for instance, the output of the processing that is returned to the GRB could contain malicious code that might bring down the entire grid. Thus, dedicated DRMs are required to avoid using publicly shared resources that could be potential points of security failure.

The single biggest disadvantage of this model is the requirement of DRMs to be dedicated to the computational grid. Practical constraints like money and space impose restrictions on the number of dedicated nodes procurable by the owners of a grid. In fact, it is easier to establish a grid amongst publicly shared DRMs, which is also more economically viable than having dedicated DRMs. This has given rise to another model of the computational grid that is prevalent today (a sample implementation being the Search for Extra-Terrestrial Intelligence: SETI@Home project [13]). It is common for an implementation of such a model to have the Internet as the connecting network. Here, the GRBs rely on DRMs that are not dedicated and are used by people not associated with the grid but which "donate" unused computing power to the grid. In this case, the same security considerations that arise in the previous model of the grid reappear. However, owing to the public nature of the underlying network connecting the grid DRMs, there are additional security issues that are to be dealt with.

In this model of the computational grid, a DRM of the grid has a 'stub' program that controls the procurement of new sets of data from the GRB, processes these sets at the DRM and transmits information back to the GRB. The key reason for a security breach in this scenario could be the use of a false 'stub' program at the node. These false 'stub's' could have been distributed over the World Wide Web, for example, by miscreants and hackers, and could have been unknowingly downloaded and installed at the DRM by its owner. This and more such issues form one aspect of security that is to be considered by users before they join a computational grid.

The other side of the coin is the security failures at the DRM that could affect the grid. Most of these failures result from loss of integrity on the part of the DRM either knowingly or unknowingly. That is, a particular DRM of the grid could be executing programs that disrupt the grid. Alternatively, due to the transparent nature of the Internet, computer viruses, for example, could infest a particular DRM and these could easily control the 'stub' program of the grid. Then, false and disruptive data could be relayed back to the grid. In fact, these events could happen without even the user of that DRM becoming aware of the security breach. It is this aspect of security that we discuss next. Note however that these issues are deemed not to exist in the earlier model where the nodes are dedicated. The SETI@Home project, perhaps the most well known example of large-scale distributed computing, has already experienced data integrity woes due to unknown and untrusted entities tampering with the computation process [21]. In order to maintain the data integrity of the Grid, we must prevent or mitigate such attacks.

As a first step, towards preventing corruption of the 'stub' program and/or the data of the grid, the user at a DRM should be prompted to install all the components of a grid program including the 'stub' and the raw data in a restricted segment of the physical memory. Either restriction could be enforced in a hardwired manner or software restrictions could be used. For example, the user could define a modified form of Access Control Lists (ACL) for the 'stub' program where only users/programs that have permissions to access that memory region are allowed access. This achieves the purpose of restricting the region over which the 'stub' program has access thus minimizing the region in memory that could potentially be affected by a false 'stub' installation as well.

Another way of protecting the grid from corrupted data/'stub' programs is to enable detection of corruption of these elements. Once an unauthorized change in data and/or the 'stub' program is detected, that particular DRM could be barred from logically connecting to the grid. A possible way of providing this could be as follows. A 'listener' program could be implemented and executed independent of the 'stub' program. The role of the 'listener' is to listen to changes happening to the program components of the grid. For example, in the case of the DRM operating with the Unix OS, the 'listener' could be monitoring the inode of the 'stub' program. An inode of a program stores meta-information pertaining to the program including the date of last update or access. Whenever a change is made to the data or a program, the 'listener' detects a change in the inode. By checking in the list of processes currently active, the 'listener' could verify to see that the 'stub' program is currently active. If it is not, then the change can be concluded to be unauthorized. Now, the 'listener' could alert the user of that node about this eventuality and prompt a relay of a message to the base station. Implementation-wise, the 'listener' program is small as only a check on the inode is required to be performed. However, some computational cycles are expended in keeping the 'listener' program active. Thus, this method could be implemented if a marginal increase in computing power usage is tolerable. A possible realization in the future could be a quantum cryptographical method that helps in instant detection of any change inflicted on data. This would provide a foolproof way of detection. Note that there exists no foolproof way of communicating the detection to the GRB as all such endeavors involve the human user. This brings us to the final issue that we discuss in this section.

A singular point of reliance of the methods illustrated above is the integrity of the *user* of the DRM. The user's complete cooperation is required for any kind of security to be implemented. No amount of security solutions can prevent a determined hacker from using his DRM as a breach point into the grid. Unfortunately, it has been the humans that are the weak links in security systems, and in many cases, it can be said that the strength of a security solution is inversely proportional to the number of humans involved in the functioning of the security system.

## IV. Conclusions

In this paper we have discussed ways to make a model of the computational grid more secure and deliver high performance simultaneously. We have suggested the use of various schemes for process authentication as well as for secure data transfer. The emphasis was on using encryption only when absolutely necessary and methods like Winnowing and Chaffing helped in this context. Another widely used model of the computational grid was also presented and some associated security issues were identified. Possible solutions and ideas were also provided. Finally, let us note that use of quantum cryptography in this area could be a big boon if efficient methods become realizable.